\documentclass[english,preprint2]{aastex}
\usepackage[T1]{fontenc}
\usepackage[latin9]{inputenc}
\usepackage{array}
\usepackage{rotating}
\usepackage{units}
\usepackage{textcomp}
\usepackage{amsbsy}
\usepackage{amstext}
\usepackage{graphicx}
\usepackage{url}
\usepackage{natbib}
\usepackage{color}

\makeatletter

\newcommand{\lyxmathsym}[1]{\ifmmode\begingroup\def\b@ld{bold}
  \text{\ifx\math@version\b@ld\bfseries\fi#1}\endgroup\else#1\fi}
\newcommand{\noun}[1]{\textsc{#1}}

\providecommand{\tabularnewline}{\\}

\usepackage{babel}
\begin{document}

\title{The Puzzling Negative Orbit-Period Derivative of the Low-Mass X-Ray Binary 4U 1820-30 in NGC 6624}

\author{M. Peuten\altaffilmark{1}, M. Brockamp\altaffilmark{1,2}, A.H.W. K\"upper\altaffilmark{3,4}, P. Kroupa\altaffilmark{5}}
\altaffiltext{1}{Argelander-Institut f\"ur Astronomie, Universit\"at Bonn, Auf dem H\"ugel 71, 53121 Bonn, Germany}
\altaffiltext{2}{Member of the International Max Planck Research School (IMPRS) for Astronomy and Astrophysics at the University of Bonn and Cologne}
\altaffiltext{3}{Department of Astronomy, Columbia University, 550 West 120th Street, New York, NY 10027, USA}
\altaffiltext{4}{Hubble Fellow}
\altaffiltext{5}{Helmholtz-Institut f\"ur Strahlen- und Kernphysik (HISKP), University of Bonn, Nussallee 14-16, 53115 Bonn, Germany}
\email{\mbox{mpeuten@mpeuten.de} (MP), \mbox{brockamp@astro.uni-bonn.de} (MB), \mbox{akuepper@astro.columbia.edu} (AHWK), \mbox{pavel@astro.uni-bonn.de} (PK)}

\begin{abstract}
4U 1820-30 is a low-mass X-ray binary near the center of the globular cluster 
NGC 6624 consisting of, at least, one neutron star and one helium white dwarf. 
Analyzing 16 years of data from the Rossi X-ray Timing Explorer (RXTE) allows 
us to measure its orbital period and its time derivative with unprecedented 
accuracy to be $P=\unit[685.01197\pm0.00003]{s}$ and 
\textbf{$\dot{P}/P=\unit[-5.3\pm0.3\times10^{-8}]{yr^{-1}}$}. Hence, we confirm 
that the period derivative is significantly negative at the $>17\sigma$ level, 
contrary to theoretical expectations for an isolated X-ray binary. We discuss
possible scenarios that could explain this discrepancy, and conclude that the 
center of NGC 6624 most likely contains large amounts of non-luminous matter 
such as dark remnants.  We also discuss the possibility of an IMBH inside NGC 
6624, or that a dark remnant close to 4U 1820-30 causes the observed shift.
\end{abstract}


\section{Introduction}

The low-mass X-ray binary (LMXB) 4U 1820-30 is the most prominent X-ray source 
in the globular cluster NGC 6624. The LMXB shows a complex variability, it has 
a 171 day periodicity with a luminosity variation of about a factor of three 
(\citealp{1984ApJ...284L..17P}). Moreover, it is the first source where X-ray 
Type I bursts have been observed (\citealp{1976ApJ...205L.127G}). These X-ray 
bursts are only seen when 4U 1820-30 is in low state. Besides the Type I bursts 
also one superburst was detected so far (\citealp{2002ApJ...566.1045S}). 
Another feature of this atoll source is that it exhibits low frequency 
(\citealp{1987ApJ...315L..49S}) and kilohertz (\citealp{1997ApJ...483L.119S}) 
quasi-periodic oscillations (QPOs). 

The feature of 4U 1820-30 which is relevant for this paper is the 685 s (11 
min) variability (\citealp{1987ApJ...312L..17S}), with a peak-to-peak 
modulation amplitude of about $\unit[\left(1.95\pm0.03\right)]{\%}$ in the 
X-ray regime. This period is generally believed to be caused by the orbital 
motion of the secondary white dwarf around the more massive neutron star 
(\citealp{1987ApJ...322..842R}). It is also observed in the ultraviolet 
(\citealp{1997ApJ...482L..69A}) with a peak-to-peak amplitude of 
$\unit[16]{\%}$. Further measurements (\citealp{1989PASJ...41..591S, 
1991ApJ...374..291T, 1993MNRAS.260..686V, 1993A&A...279L..21V}) of the 
$\unit[685]{s}$ periodicity showed that its period derivative seems to be 
negative. The last measured value by \citet{2001ApJ...563..934C} 
$\dot{P}/P=\unit[\left(-3.47\pm1.48\right)\times10^{-8}]{yr^{-1}}$ is negative, 
but due to the large uncertainty a positive value cannot be excluded at high 
confidence.

NGC 6624 is a bulge globular cluster, approximately $\unit[1.2]{kpc}$
(\citealp{1996AJ....112.1487H}) away from the Galactic Center and 
$\unit[7.9]{kpc}$ (\citealp{1996AJ....112.1487H}) away from Earth. According to
the Harris catalog, NGC 6624 has a heliocentric radial velocity of 
$\left\langle \nu_{r}\right\rangle =\unit[53.9\pm0.6]{km\, s^{-1}}$ and a 
radial velocity dispersion of $\sigma_{\nu}\approx\unit[8.6]{km\, s^{-1}}$.
Like most of its bugle/disk counterparts, the cluster is metal-rich with an 
iron abundance of $[Fe/H]=-0.69\pm0.02$ (\citealp{2011MNRAS.414.2690V}). But 
NGC 6624 is very compact, its half-light radius is 
$r_{h}=\unit[49"2\approx1.88]{pc}$ (\citealp{1996AJ....112.1487H}), and it 
appears to be core-collapsed with a radial density profile that has a power-law 
cusp with a slope of about $-0.8$ to $-0.3$ (\citealp{1995AJ....109..639S, 
2006AJ....132..447N}). Early measurements gave a mass density of the core of 
about $\rho_{c}\approx\unit[1.1\times10^{5}]{M_{\odot}\, pc^{-3}}$ 
(\citealp{1978ApJ...224...39C}) and a luminosity density of about 
$\log_{10}(\rho_{0})=\unit[5.25]{L_{\odot}\,pc^{-3}}$ 
(\citealp{2002A&A...391..945P}). 

One of the first theoretical descriptions of 4U 1820-30 was done by 
\citet{1987ApJ...322..842R} who described the LMXB as consisting of a neutron 
star and a degenerate almost pure, or entirely pure, helium dwarf. According 
to this interpretation, the secondary fills its Roche lobe and thus transfers 
matter to the accretion disc of the neutron star. This theory was later 
extended by \citet{1993ApJ...413L.121A} who showed that the high ultraviolet 
luminosity could be explained by a reprocessing of the emitted X-ray photons. 
One of the observations which had not been explained by 
\citet{1987ApJ...322..842R}, the 171 days period, was later explained by 
\citet{2012ApJ...747....4P}, who took up the idea of a third bound star. This 
scenario had already been suggested by \citet{1988IAUS..126..347G} and again by
\citet{2001ApJ...563..934C}. \citet{2012ApJ...747....4P} calculated the 
properties needed for the additional star to modulate the accretion rate via 
the Kozai mechanism.

To date, many of the predictions made by \citet{1987ApJ...322..842R} have been 
confirmed by observations, but there is one important prediction that is not 
met yet: \citet{1987ApJ...322..842R} predict that the period derivative of the 
685 s period must be positive, as the secondary is losing mass to the neutron 
star. The X-rays which are detected from 4U 1820-30 are a direct consequence of 
this process. That is, the orbital period of the secondary should be increasing
with time. \citet{1987ApJ...322..842R} further predict a value of around 
$\dot{P}/P\approx\unit[8.8\times10^{-8}]{yr^{-1}}$ for the period derivative. 
When calculating this value, the authors neglected some effects (see 
\citet{2012ApJ...747....4P} for a discussion) which could lower the intrinsic 
period derivative, but these effects should not change the sign of $\dot{P}/P$. 
Thus, the period derivative should nevertheless be positive, but the observed 
value seems to be negative. However, as already stated, the latest measurement 
does not exclude at a high significance that the true value could be positive.

The motivation of the present paper is to confirm the existence of a negative 
period derivative with higher statistical significance than before. To achieve 
this goal we analyzed sixteen years of Rossi X-ray Timing Explorer (RXTE) data, 
which had not been analyzed in this respect so far. We are now able to show 
with an updated value of 
$\dot{P}/P=\unit[\left(-5.3\pm0.3\right)\times10^{-8}]{yr^{-1}}$ that, contrary 
to the theoretical predictions, the period derivative is truly negative at high 
confidence. It is smaller than the result from \citet{2001ApJ...563..934C} but 
equal to the value that \citet{1993A&A...279L..21V} had measured before. 

In Section \ref{sec:Data,-Data-Reduction-Analysis}, we give an overview about
the observations. In the same section, we also present a detailed description 
of the different steps involved in the reduction of the RXTE/PCA data. The 
subsequent Section \ref{sec:Results} describes our analysis of the results from 
the reduction, graphically presented in Fig.~\ref{fig:Phi-Plot}, and our 
calculation of the period derivative as being 
$\dot{P}/P=\unit[\left(-5.3\pm0.3\right)\times10^{-8}]{yr^{-1}}$.
In Section \ref{sec:Interpretation}, we discuss some possible explanations for 
the difference between theoretical predictions and the value we measured: We 
conclude that this difference is primarily generated by gravitational 
acceleration within the cluster. We therefore discuss in this section the 
probable sources of such a gravitational acceleration such as an Intermediate 
Mass Black Hole (IMBH), a high central concentration of dark remnants or a 
close flyby of a single dark remnant such as a stellar-mass black hole or 
neutron star. A summary of our conclusions can be found in Section 
\ref{sec:Conclusion}.

\section{Data Reduction and Analysis\label{sec:Data,-Data-Reduction-Analysis}}

\subsection{Available Datasets}

The Rossi X-ray Timing Explorer (RXTE) satellite was launched in December 1995 
and the mission lasted for 16 years until January 2012. The primary aim of RXTE 
was to analyze the X-ray universe with a high temporal resolution and a medium 
energy resolution. The RXTE consists of three instruments. We here use data 
from the Proportional Counter Array (PCA) \citep{1996SPIE.2808...59J,
2006ApJS..163..401J}. The PCA has a high temporal resolution of 
$\unit[1]{\mu s}$ over an energy range of $\unit[2-60]{keV}$ with an energy 
resolution of $\unit[18]{\%}$ (at $\unit[6]{keV}$). 

Parts of the 1996 and 1997 data were already analyzed by 
\citet{2001ApJ...563..934C}, but for the sake of comparability and because of 
the fact that, in the meantime, newer reduction software and calibration data 
became available \citep{2012ApJ...757..159S}, we reevaluated them in this 
paper. We adopted the analysis approach from \citet{2001ApJ...563..934C} as 
they worked with the same satellite and the same target. 

\begin{figure*}
\includegraphics[width=1\textwidth]{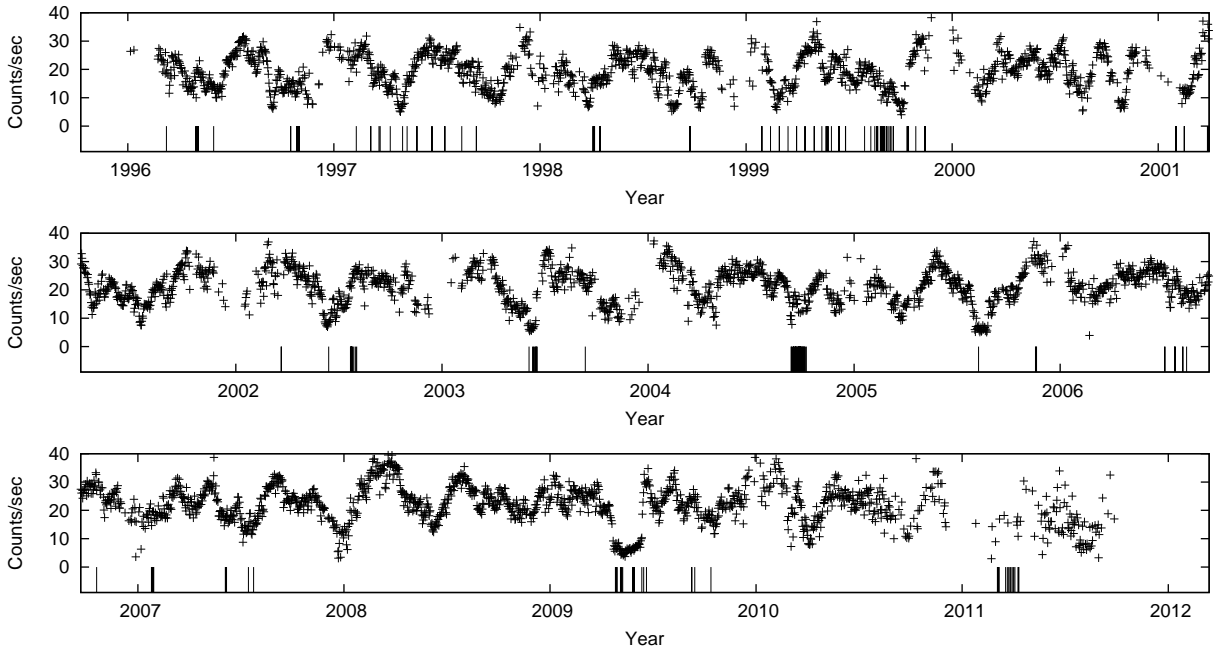}
\caption{Plot showing the distribution of the observations over the whole 
lifetime of the RXTE mission. The black crosses are the ASM counts of 4U 
1820-30 while each black bar represents a PCA observation of 4U 1820-30. 
\label{fig:ASMvsPCU}}
\end{figure*}

For our analysis we used all available observations of 4U 1820-30 made by the 
PCA. These 261 observations were made throughout the whole lifetime of the RXTE 
mission, which includes data from 1996 to 2011. Fig.~\ref{fig:ASMvsPCU} shows a 
plot of the measurements of 4U 1820-30 from the All-Sky Monitor (ASM, 
\citealt{1996ApJ...469L..33L}), which was also on board the RXTE mission, 
together with bars indicating the time when observations were made with the PCA 
instrument. It is apparent that almost half of all observations were made in 
the first four years of the mission. In the remaining twelve years, the PCA 
instrument observed 4U 1820-30 mainly when it was in low state, in order to 
catch some of the X-ray Type I bursts. As previous publications 
(for example \citealt{1988ApJ...324..851M}) showed no correlation between the 
relative amplitude of the 685 sec periodicity and the overall X-ray luminosity, 
the results from these years are subject to a lowered statistic.

From the available PCA modes we analyzed the \noun{Standard-2} data as it is 
available from all observations. The PCA \noun{Standard-2} data has a time 
resolution of $\unit[16]{s}$ and an energy resolution of 129 bit covering the 
whole PCA energy range from $2$ to $\unit[60]{keV}$. The PCA instrument 
consists of five equal Proportional Counter Units (PCU). Throughout the mission
these five units performed quite differently. For this project we decided to 
only use data from PCU2 which was not only the most stable one but also in many
observations the only observing one. There were only two observations 
(Observations 40017-01-71-00 and 40017-01-17-01) where PCU2 was deactivated due
to an earlier breakdown. We therefore decided to use data from PCU0 for these 
time ranges. Until May 2000, when PCU0 was hit by a micrometeorite, it 
performed as well as PCU2. 

To exclude energy specific systematics, we performed the complete analysis in 
different energy ranges ($\unit[1-5]{keV}$, $\unit[5-8]{keV}$, 
$\unit[8-20]{keV}$, $\unit[20-60]{keV}$ and full PCA energy range), but found, 
as expected from previous observations, no correlation between the results and 
the selected energy range. For this reason we present here the analysis of the 
complete energy band ($\unit[2-60]{keV}$) of the PCA instrument. In the 
reduction and the following analysis we used the HEASOFT V6.12 (as of March 12,
2012) \footnote{\url{https://heasarc.gsfc.nasa.gov/lheasoft/}} software. 

\subsection{Data Reduction}

We started the reduction by filtering the raw observational data for ``bad 
times``. Bad times are periods with some sort of interference, like Earth 
occultation, passage of the South Atlantic Anomaly or any other events which 
can interfere with the detector. Next, the resulting light curves were 
inspected for anomalies such as X-ray bursts. Since these bursts interfere with
the 685 sec signal, they had to be removed from the data. For the one 
observation (Observation 30057-01-04-08G) where the superburst was observed, we 
decided to remove the whole observation because the superburst almost dominates
the entire observation. All in all we found eighteen X-ray bursts in the RXTE 
data, including the superburst. After identifying all unusable time periods, we 
reran the filtering step, but this time by leaving out inapplicable data. In 
the next step, we generated the background for the different observations 
(see \citealp{2006ApJS..163..401J} for a in depth description) and then we 
subtracted it from the associated observations. Afterwards, the data was 
corrected for barycentric arrival times ($\unit{BJD_{\scriptsize{\mbox{BDT}}}}$).
This is required to have a location-independent time stamp on all data.

\subsection{Data Analysis}

\begin{figure}
\includegraphics[width=1\columnwidth]{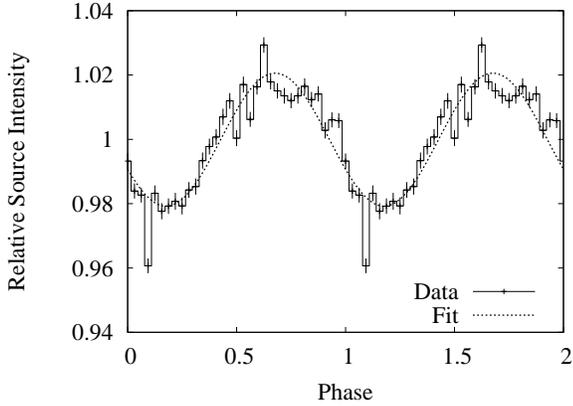}
\caption{Folded light curve from Observation 40017-01-05-00: The solid black 
histogram is the folded light curve with $1\sigma$ errors. The dotted black 
curve is the best fit sine function. One can clearly see the sine-like behavior.
The obtained phase value is: $0.68\pm0.01$. \label{fig:Folded-Light-good} }
\end{figure}

\begin{figure}
\includegraphics[width=1\columnwidth]{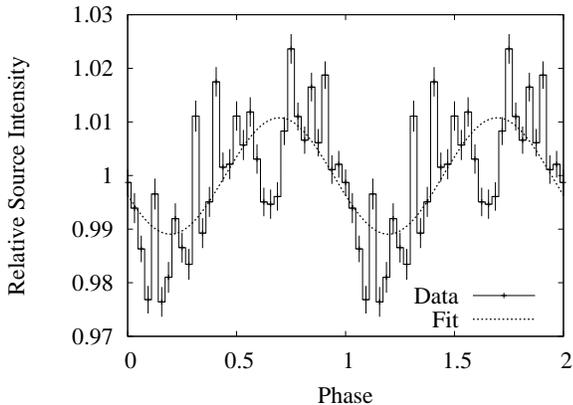}
\caption{Folded light curve from Observation 40017-01-04-00: The solid black 
histogram is the folded light curve with $1\sigma$ errors. The dotted black 
curve is the best fit sine function. Besides the sine-like behavior, one can 
clearly see the second dip. The obtained phase value is: $0.70\pm0.04$
\label{fig:-Folded-Light-duble-dip}}
\end{figure}

The next step of the analysis consists of epoch folding the individual light 
curves. But when comparing the length of the different observations, we found 
that there are many short observations, i.e. shorter than one RXTE orbit 
$\approx\unit[90]{min}$. There were even ten light curves which not even 
covered the whole 685 sec period. However, many of these short observations 
were taken on the same day. We therefore decided to join together the light 
curves of observations being no further away from each other than 24h. This 
improves the results from the epoch folding method as this technique needs more 
than one complete period of 685 sec to give an improved result. The averaged 
observation length of all resulting 154 light curves was 2.6 h, of which none 
were shorter than the 685 sec period of the LMXB. Before folding the light 
curves they got rebinned. Following \citet{2001ApJ...563..934C}, we chose 32 
phase bins, as varying this value did not give any improvements. Afterwards, 
these light curves were finally folded with the ephemeris 
by \citet{1991ApJ...374..291T}:
\begin{equation}
\begin{array}{cc}
T_{N}^{\scriptsize{\mbox{max}}}= & \unit{HJD_{\scriptsize{\mbox{UTC}}}}2442803.63544\\
 & +\left(\frac{\unit[685.0118]{s}}{\unit[86400]{s}}\right)\times N
\end{array}
\end{equation}
where $T_{N}^{\scriptsize{\mbox{max}}}$ is the time for the maximum of the 
$N$-th cycle. Since our data had $\unit{BJD_{\scriptsize{\mbox{BDT}}}}$ time 
stamps, we had to convert the above formula (see \citealp{2010PASP..122..935E})
to:
\begin{equation}
\begin{array}{cc}
T_{N}^{\scriptsize{\mbox{max}}}= & \unit{BJD_{\scriptsize{\mbox{TDB}}}}2442803.63600\\
 & +\left(\frac{\unit[685.0118]{s}}{\unit[86400]{s}}\right)\times N
\end{array}\label{eq:Tan_BJD}
\end{equation}
In Fig.~\ref{fig:Folded-Light-good} we present one of the resulting folded 
light curves. The sine-like behavior of the 685 sec period, as also found by 
all previous analysis, is clearly evident. In some of the folded light curves 
we also detected a second minimum as shown in 
Fig.~\ref{fig:-Folded-Light-duble-dip}. This second minimum was first observed 
by \citet{1988ApJ...324..851M} and later by \citet{1989PASJ...41..591S}, 
\citet{1991ApJ...374..291T} and \citet{1993MNRAS.260..686V,1993A&A...279L..21V}.
This minimum seems to appear randomly with changing positions and amplitude. 
Moreover, as the observation times are randomly distributed, we performed no 
extensive analysis of this phenomenon. 

In order to determine the phase of the maximum we approximated a sine-function 
to the folded light curves. The results are also shown in 
Fig.~\ref{fig:Folded-Light-good} and \ref{fig:-Folded-Light-duble-dip}. For one 
folded light curve (from Observation 90027-01-01-03, September 13, 2004), a 
sine curve approximation failed and this data point was therefore removed. As 
the phase is periodic we took care that the resulting values were as continuous
to each other and to previous results as possible. This is a reasonable 
assumption, as there were no indications against it. The sine fit allows us to 
also determine the average modulation amplitude (peak to peak). We measured
$\unit[\left(1.95\pm0.03\right)]{\%}$.

\begin{figure*}
\includegraphics[width=1\textwidth]{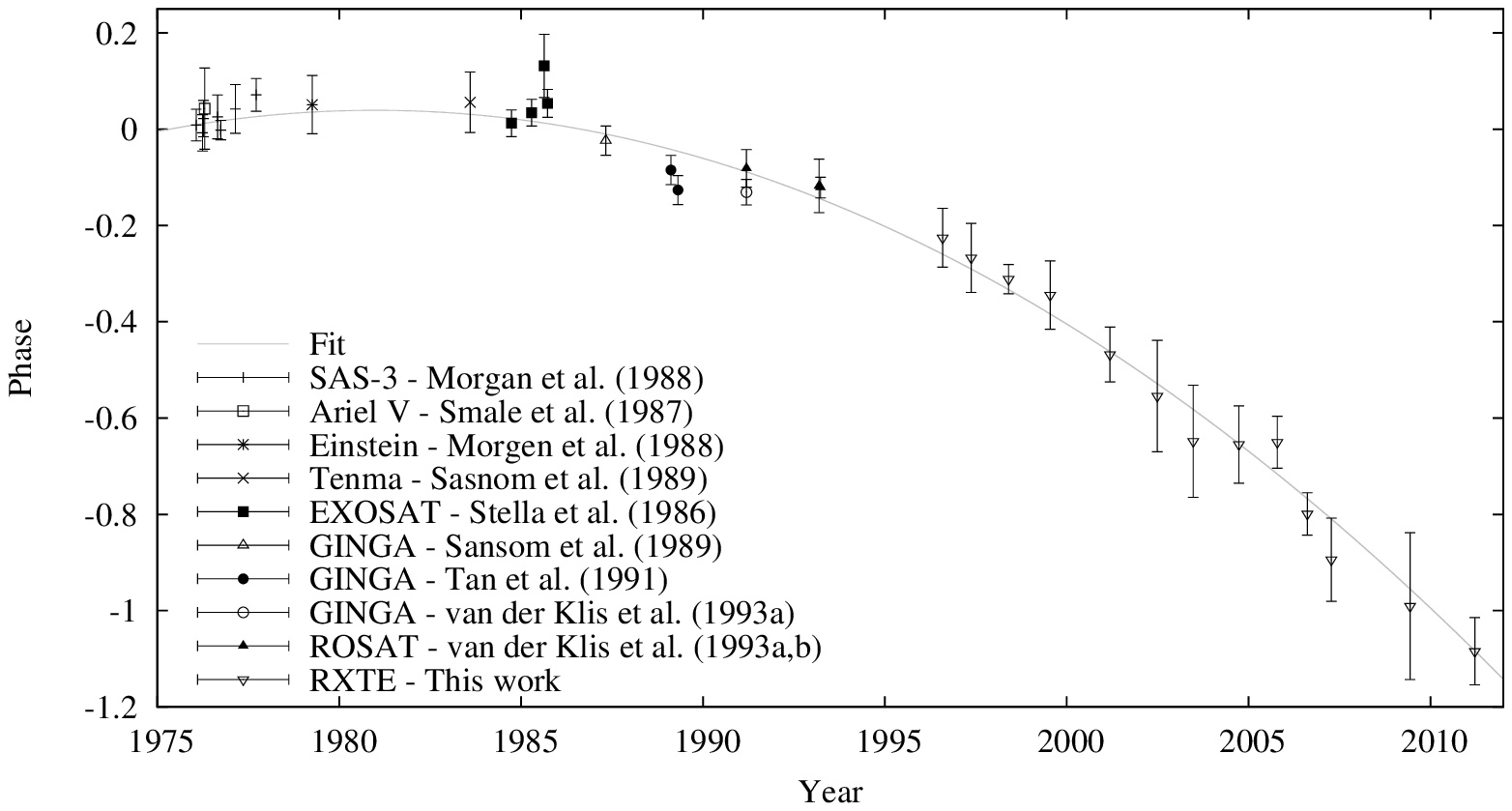}
\caption{Phases of the 685 sec modulation of 4U 1820-30 from X-ray satellite
observations from 1976 to 2011. The solid line represents the best quadratic 
fit from which the new value for the period derivative is determined.
\label{fig:Phi-Plot}}
\end{figure*}

Following \citet{2001ApJ...563..934C} the obtained phases from the same year 
were joined together. This was done by weighted averaging the individual phase 
values from the same year. To deal with the scattering of individual phases 
around the mean value, we quadratically added the mean phase jitter, which is 
the weighted standard deviation of the phase values, to the mean phase error. 
The results from this last step are given in Table \ref{tab:PHI-OLD} and are 
presented in Fig.~\ref{fig:Phi-Plot}. Besides our results, we also listed and 
plotted the results from previous publications as listed by 
\citet{1993A&A...279L..21V,1993MNRAS.260..686V}. These datasets come from the 
following X-ray missions: SAS-3 \citep{1988ApJ...324..851M}, Ariel V 
\citep{1987MNRAS.225P...7S}, Einstein \citep{1988ApJ...324..851M}, Tenma 
\citep{1989PASJ...41..591S}, EXOSAT \citep{1986BAAS...18.1048S}, GINGA 
(\citealp{1989PASJ...41..591S}, \citealp{1991ApJ...374..291T} and 
\citealp{1993MNRAS.260..686V}) and ROSAT \citep{1993MNRAS.260..686V,
1993A&A...279L..21V}. In order to combine our results with those datasets, we 
first had to convert their time stamps into 
$\unit{BJD_{\scriptsize{\mbox{TDB}}}}$. With the reduced and analyzed RXTE data 
at hand, together with the data from previous missions adapted to our 
requirements, we can now determine the proper value of $\dot{P}/P$. This will 
be the topic of the next Section~\ref{sec:Results}.

\begin{figure}
\includegraphics[width=1\columnwidth]{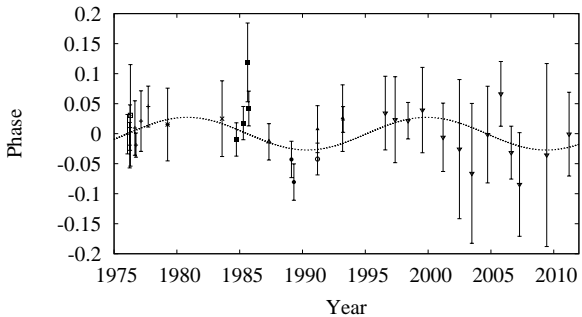}
\caption{Differences (residuals) between the measured phase shifts and our 
quadratic solution. There seems to be a $\unit[19\pm2]{yr}$ modulation 
($\chi_{\scriptsize{\mbox{red}}}^{2}=0.70$) which is roughly two times the 
$\approx\unit[8.5\pm0.2]{yr}$ sinusoidal curve as proposed by 
\citet{1991ApJ...374..291T}.\label{fig:Phi-Plot-Res}}
\end{figure}

\begin{table*}
\begin{center}
\begin{tabular}{@{}cccc}
\hline 
Arrival time of Maximum {[}$\unit{BJD_{\scriptsize{\mbox{TDB}}}}${]} & Phase & Satellite & Reference\tabularnewline
\hline 
\hline 
$2442803.64$ & $0.01\pm0.03$ & SAS-3 & 1\tabularnewline

$2442866.86$ & $-0.01\pm0.04$ & SAS-3 & 1\tabularnewline

$2442876.81$ & $0.02\pm0.04$ & SAS-3 & 1\tabularnewline

$2442889.19$ & $0.04\pm0.09$ & Ariel V & 2\tabularnewline

$2443018.64$ & $0.03\pm0.05$ & SAS-3 & 1\tabularnewline

$2443050.95$ & $0.00\pm0.02$ & SAS-3 & 1\tabularnewline

$2443196.93$ & $0.04\pm0.05$ & SAS-3 & 1\tabularnewline

$2443406.11$ & $0.07\pm0.03$ & SAS-3 & 1\tabularnewline

$2443969.20$ & $0.05\pm0.06$ & Einstein & 1\tabularnewline

$2445554.11$ & $0.06\pm0.06$ & Tenma & 3 \tabularnewline

$2445969.88$ & $0.01\pm0.03$ & EXOSAT & 4 \tabularnewline

$2446172.22$ & $0.03\pm0.03$ & EXOSAT & 4\tabularnewline

$2446297.41$ & $0.13\pm0.07$ & EXOSAT & 4\tabularnewline

$2446331.57$ & $0.05\pm0.03$ & EXOSAT & 4\tabularnewline

$2446916.51$ & $-0.02\pm0.03$ & GINGA & 3\tabularnewline
 
$2447569.84$ & $-0.08\pm0.03$ & GINGA & 5\tabularnewline

$2447641.36$ & $-0.13\pm0.03$ & GINGA & 5\tabularnewline

$2448327.96$ & $-0.13\pm0.03$ & GINGA & 6\tabularnewline

$2448328.42$ & $-0.08\pm0.040$ & ROSAT & 6 \tabularnewline

$2449057.60$ & $-0.12\pm0.06$ & ROSAT & 7\tabularnewline

$2449071.40$ & $-0.12\pm0.02$ & ROSAT & 7\tabularnewline
\hline 
$2450299.23$ & $-0.23\pm0.06$ & RXTE & This work \tabularnewline

$2450584.90$ & $-0.27\pm0.07$ & RXTE & \textquotedbl \tabularnewline

$2450960.23$ & $-0.31\pm0.03$ & RXTE & \textquotedbl \tabularnewline

$2451378.59$ & $-0.34\pm0.07$ & RXTE & \textquotedbl \tabularnewline

$2451978.78$ & $-0.47\pm0.06$ & RXTE & \textquotedbl \tabularnewline

$2452453.69$ & $-0.6\pm0.1$ & RXTE & \textquotedbl \tabularnewline

$2452815.24$ & $-0.6\pm0.1$ & RXTE & \textquotedbl \tabularnewline

$2453273.76$ & $-0.65\pm0.08$ & RXTE & \textquotedbl \tabularnewline

$2453659.48$ & $-0.65\pm0.05$ & RXTE & \textquotedbl \tabularnewline

$2453961.10$ & $-0.80\pm0.04$ & RXTE & \textquotedbl \tabularnewline

$2454201.79$ & $-0.89\pm0.07$ & RXTE & \textquotedbl \tabularnewline

$2454991.88$ & $-1.0\pm0.2$ & RXTE & \textquotedbl \tabularnewline

 $2455643.48$ & $-1.08\pm0.07$ & RXTE & \textquotedbl \tabularnewline
\hline 
\end{tabular}
\caption{Results from previous missions as listed by 
\citet{1993A&A...279L..21V,1993MNRAS.260..686V} converted to 
$\unit{BJD_{\scriptsize{\mbox{TDB}}}}$ time stamps together with results from 
our analysis of the RXTE data from 4U 1820-30. The error values for the arrival
times are below $\unit[10^{-3}]{days}$ and are therefore not listed. Note: The 
resulting RXTE phase values are chosen to be continuous with values from the 
previous publications. \textbf{References:} (1) \citet{1988ApJ...324..851M}; (2)
\citet{1987MNRAS.225P...7S}; (3) \citet{1989PASJ...41..591S}; (4) 
\citet{1987ApJ...312L..17S}; (5) \citet{1991ApJ...374..291T}; (6) 
\citet{1993MNRAS.260..686V}; (7) \citet{1993A&A...279L..21V}.
\label{tab:PHI-OLD}}
\end{center}
\end{table*}

\section{Results\label{sec:Results}}

In Fig.~\ref{fig:Phi-Plot} one sees how the phase shift evolves over time. The 
data were folded with the linear ephemeris as given by 
\citet{1991ApJ...374..291T} (see also Eq.~\ref{eq:Tan_BJD}). If the true 
ephemeris would be really linear, one would expect the phase shift to be 
constant. This is apparently not the case: the phase shift is decreasing. This 
behavior can be approximated very well with the true ephemeris having a 
quadratic time dependence. The observed phase shift in Fig.~\ref{fig:Phi-Plot} 
is therefore the difference between the two ephemerides. Following 
\citet{2001ApJ...563..934C}, we subtracted Eq.~\ref{eq:Tan_BJD} from a 
quadratic ephemeris to arrive at
\begin{equation}
\Phi(t)=\Phi_{0}+\frac{\Delta P}{\left(P_{fold}\right)^{2}}t+\frac{1}{2}\frac{\dot{P}}{\left(P_{fold}\right)^{2}}t^{2}.\label{eq:Phi-vs-t}
\end{equation}
Here $\Phi(t)$ is the measured shift over the time 
$t=T_{N}^{\scriptsize{\mbox{max}}}-T_{\scriptsize{\mbox{fold}}}$, where 
$T_{N}^{\scriptsize{\mbox{max}}}$ is the time of maximum of the $N$-th cycle and
$T_{\scriptsize{\mbox{fold}}}=\unit{BJD_{\scriptsize{\mbox{TDB}}}}2442803.63600$
corresponds to the offset in the ephemeris from \citet{1991ApJ...374..291T} and 
$P_{\scriptsize{\mbox{fold}}}=\unit[(685.0118/86400)]{days}$ is the period of 
the same ephemeris. The phase offset is given by 
$\Phi_{0}=\left(T_{0}-T_{\scriptsize{\mbox{fold}}}\right)/P_{\scriptsize{\mbox{fold}}}$ 
and the period difference is $\Delta P=P_{0}-P_{\scriptsize{\mbox{fold}}}$. 
$T_{0}$, $P_{0}$ and $\dot{P}$ are the true offset, period and period 
derivative of the real ephemeris of 4U 1820-30, which we want to determine. To 
finally calculate them we approximate Eq.~\ref{eq:Phi-vs-t} to the dataset as 
given in Table \ref{tab:PHI-OLD} (see Fig.~\ref{fig:Phi-Plot} and 
Fig.~\ref{fig:Phi-Plot-Res} for the residuals).
 
As shown in Fig.~\ref{fig:Phi-Plot-Res} the residuals seem to be compatible 
($\chi_{\scriptsize{\mbox{red}}}^{2}=0.70$) to a sinusoidal modulation having a
$\unit[19\pm2]{yr}$ period. This is approximately two times 
the $\approx\unit[8.5\pm0.2]{yr}$ sinusoidal period which was solely identified
by \cite{1991ApJ...374..291T} and interpreted as an artifact subject to the 
sparse time coverage of the data. However, our dataset covers more than twice 
the time interval than theirs and it indicates an intrinsic origin of the 
source rather than an observational bias. We postpone the interpretation of 
Fig.~\ref{fig:Phi-Plot-Res} to future studies.

This yields a period of $P=\unit[\left(685.01197\pm0.00003\right)]{s}$ and
derivative of 
$\dot{P}=\unit[\left(-1.15\pm0.06\right)\times10^{-12}]{s\, s^{-1}}$ (at
$T_{0} = \unit{BJD_{\scriptsize{\mbox{TDB}}}}2442803.63608\pm0.000094$).
This translates into 
$\dot{P}/P=\unit[\left(-5.3\pm0.3\right)\times10^{-8}]{yr^{-1}}$. The reduced 
$\chi^{2}$ value centers close to one, i.e 
$\chi_{\scriptsize{\mbox{red}}}^{2}=0.88$. The $95\%$ confidence interval for 
the period derivative is 
$\unit[-1.279\times10^{-12}]{s\, s^{-1}} \leq \dot{P} \leq \unit[-1.019\times10^{-12}]{s\, s^{-1}}$
, respectively 
$\unit[-5.89\times10^{-8}]{yr^{-1}} \leq \dot{P}/P \leq \unit[-4.69\times10^{-8}]{yr^{-1}}$. 
Our $\dot{P}/P$ value is equal to the one found by \citet{1993A&A...279L..21V}
and is also consistent with the value from \citet{2001ApJ...563..934C}. 
However, we lowered the error by at least a factor of four and can rule out any
positive period derivative with a significance of $>17\sigma$. The fact that 
the decrease of the period must be faster than \citet{2001ApJ...563..934C} 
claimed was already suspected by \citet{2007MNRAS.377.1017Z} and 
\citet{2011ApJS..196....6L}. 

In order to check whether the quadratic function approximates the data more 
accurately than the linear regression 
($\chi_{\scriptsize{\mbox{red}}}^{2}=9.81$), we performed an F-test which 
passed the $99.99\%$ confidence level which we enforced. A cubic fit 
($\chi_{\scriptsize{\mbox{red}}}^{2}=0.91$) on the other hand failed the F-test
at the same $99.99\%$ confidence level. Using our RXTE data alone is not 
sufficient to determine $\dot{P}$, since the F-test for the quadratic fit 
($\chi_{\scriptsize{\mbox{red}}}^{2}=0.35$) vs. the linear fit 
($\chi_{\scriptsize{\mbox{red}}}^{2}=0.39$) fails at the $99.99\%$ confidence 
level. Still, the obtained value of 
$\dot{P}/P=\unit[\left(-4.1\pm2.7\right)\times10^{-8}]{yr^{-1}}$ is in 
agreement with the result from the combined dataset, although with an 
unacceptable large error value. 

The updated ephemeris can now be written as:
\begin{equation}
\begin{array}{cc}
T^{\scriptsize{\mbox{max}}}_{N}= & \unit{BJD_{\scriptsize{\mbox{TDB}}}}2442803.63608\pm0.000094\\
 & +\left(\frac{\unit[685.01197\pm0.00003]{s}}{\unitfrac[86400]{s}{d}}\right)\times N+\\
 & +\left(\left(-4.6\pm0.3\right)\times10^{-15}\right)\times N^{2}
\end{array}
\end{equation}

\section{Interpretation\label{sec:Interpretation}}

As already presented in the introduction, the theoretically expected value for 
the period derivative is positive and should, according to 
\citet{1987ApJ...322..842R}, have a value of 
$\dot{P}/P>8.8\times10^{-8}\ yr^{-1}$. Their arguments are based on Roche 
overflow from the secondary to the primary neutron star. 
\citet{2012ApJ...747....4P} discussed internal effects, which could also lower 
the intrinsic period derivative predicted by \citet{1987ApJ...322..842R}. 
Although these effects would decrease the value, but they would not change its
sign. Based on the fact that many other predictions made by 
\citet{1987ApJ...322..842R} are in good agreement with the measurements, it is 
reasonable to assume that the intrinsic period derivative is indeed positive 
(\citet{2012ApJ...747....4P}). It has to been investigated whether additional 
internal effects, like Applegate's Gravitational Quadrupole Coupling (GQC) 
(\citealp{1992ApJ...385..621A}) model, which successfully explains the negative
period derivative of the pulsar PSR J2051\textminus{}0827 
(\citealp{2011MNRAS.414.3134L}), can also explain the negtive period derivative.

However, our approach bases on external effects, which are also known to explain 
negative period derivatives. The most straightforward explanation for the 
negative period derivative is a Doppler shift caused by gravitational 
acceleration of 4U 1820-30 towards us. This idea was already discussed for 4U 
1820-30 by \citet{1993MNRAS.260..686V}, \citet{2001ApJ...563..934C} and 
\citet{2012ApJ...747....4P}. The theoretical idea stems from 
\citet{1992RSPTA.341...39P} who studied this effect as means to measure 
globular cluster potentials with the help of period derivatives from pulsars. 
\citet{2011MNRAS.410.2698G} applied this method convincingly in their 
Monte-Carlo modeling of the Milky Way globular cluster 47 Tucanae. We follow 
their approach here.

\citet{1992RSPTA.341...39P} found the following relation for globular clusters:
\begin{equation}
\frac{\dot{P}_{\scriptsize{\mbox{int}}}}{P}\leq\frac{\dot{P}_{\scriptsize{\mbox{obs}}}}{P}+\frac{a_{c}}{c}+\frac{a_{G}}{c}+\frac{\mu^{2}D}{c}
\end{equation}
where $\dot{P}_{\scriptsize{\mbox{int}}}$ is the intrinsic period derivative in
the rest frame, $\dot{P}_{\scriptsize{\mbox{obs}}}$ the observed period 
derivative, $a_{c}$ the acceleration due to the cluster potential, $a_{G}$ the
acceleration due to the galaxy, $\mu$ the proper motion and D the distance to 
the cluster. The last term is the Shklovskii effect 
(\citealt{1970SvA....13..562S}) which is positive but usually sub-dominant. The
value for galactic acceleration for NGC 6624 is around 
$\frac{a_{G}}{c}\approx\unit[-5.7\times10^{-11}]{yr^{-1}}$ (applying Equation 
2.3 from \citealt{1992RSPTA.341...39P}) and can therefore be neglected. 

Following \cite{1992RSPTA.341...39P}, the expected period derivative change
in a cluster potential is: 
\begin{equation}
\frac{a_c}{c} = - \frac{1}{c}\frac{GM(<r)}{r^{2}}\frac{l}{r},
\label{eq:Fig6}
\end{equation}
where $M(<r)$ is the cluster mass within the radius $r$ and $l$ is the distance
along the line-of-sight (LOS) from the source to the line which is 
perpendicular to the LOS and which passes right through the center of the 
globular cluster. If the pulsar lies behind the center, $l$ is positive and 
hence the Doppler shift is negative.

The maximal possible acceleration at a given cluster radius depends on the 
mass distribution within the cluster. For a cored mass profile with a core
radius $r_{c}$, \citet{1993ASPC...50..141P} estimated the maximal acceleration
to be
\begin{equation}
\frac{a_{c,max}}{c}\approx\frac{3\sigma(R_{\bot})^{2}}{2c\left(r_{c}^{2}+R_{\bot}^{2}\right)^{1/2}},
\end{equation}
with $\sigma(R_{\bot})$ being the line-of-sight velocity dispersion and 
$R_{\bot}$ being the projected distance. Using this method, 
\citet{2012ApJ...747....4P} calculated the cluster acceleration to be
$\frac{a_{\scriptsize{\mbox{c,max}}}}{c}=\unit[1.3\times10^{-9}]{yr^{-1}}$ at 
the assumed position of 4U 1820-30. This value is significantly too low to be 
able to explain the observed period change. But their estimate of the 
acceleration assumes NGC 6624 to have a cored density profile. However, as 
already stated, NGC 6624 appears to be in deep core collapse and therefore 
shows a central cusp. Based on Fokker-Planck models of 
\citet{1993MNRAS.260..686V}, \citet{2001ApJ...563..934C} estimated the 
gravitational acceleration on 4U 1820-30 to be
$\frac{a_{\scriptsize{\mbox{c,max}}}}{c}\approx\unit[7.9\times10^{-8}]{yr^{-1}}$. 
This value would be sufficient to explain the observed period derivative by 
assuming a negligible contribution from the internal positive period 
derivative. However, \citet{2001ApJ...563..934C} used the projected distance
$\unit[0.02]{pc}$ of 4U 1820-30 to the center of NGC 6624 which is nowadays 
considered to be outdated by more recent observations. Using HST/ACS imaging 
data, \citet{2010AJ....140.1830G} have redetermined the center of NGC 6624 to 
be slightly offset from previous HST determinations, which puts 4U 1820-30 at a
projected radius of 0.046 pc from the center. Since the modeling of the cluster
acceleration on 4U 1820-30 is very sensitive to this value, we here made a new 
modeling approach based on the latest observational data on NGC 6624.  

\subsection{Modeling of NGC 6624}

\begin{table*}
\begin{center}
\begin{tabular}{@{}cccc}
Name &R [arcsec] & R [pc]& $\dot{P}/P\ [\unit[10^{-8}]{yr^{-1}}]$ \tabularnewline
\hline 
\hline 
J1823\textminus{}3021A & $0.52$ & $0.020$ & $1.9636\pm0.0006$\tabularnewline
J1823\textminus{}3021B & $13.46$ & $0.516$ & $0.263\pm0.003$\tabularnewline
J1823\textminus{}3021C & $8.39$ & $0.152$ & $1.74\pm0.02$\tabularnewline
4U 1820\textminus{}30 & $1.30$ & $0.050$ & $-5.3\pm0.3$\tabularnewline
\hline 
\end{tabular}
\caption{List of pulsars with known period derivatives in NGC 6624. The 
corresponding values for 4U 1820-30 were also added. List compiled from data
from \citet{2012ApJ...745..109L}, \citet{2010AJ....140.1830G}, 
\citet{2011A&A...533A..33Z} and this work. See also accompanying 
Fig.~\ref{fig:a_over_c_vs_d}.\label{tab:List-of-Pulsars}}
\end{center}
\end{table*}

\begin{figure*}
\centering
\includegraphics[width=1\textwidth]{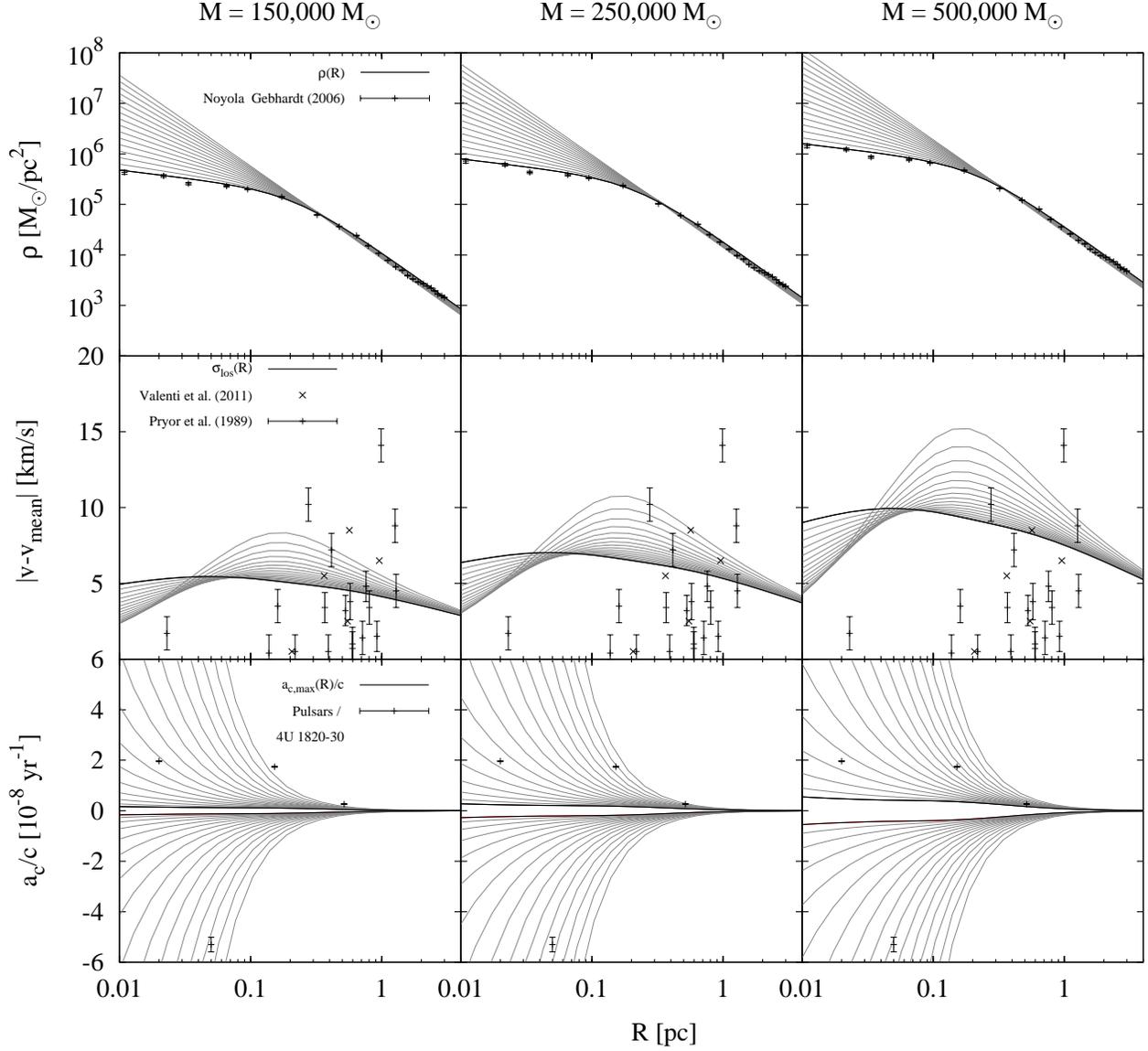}
\caption{Comparison of cluster models of $1.5\times10^5$\,M$_\odot$ (left 
column), $2.5\times10^5$\,M$_\odot$ (middle column), and  $5.0\times10^5$\,
M$_\odot$ (right column), with observational data. Top row: The density profile
of the models were generated using NGC 6624's observed surface brightness 
profile and assuming that mass follows light (thick black line) or that the 
mass density profile slope is steeper than the surface brightness slope of 0.32
towards the cluster center with slopes of 0.4 to 1.8 (gray lines). Middle row: 
Line-of-sight velocity dispersions for the models compared to 23 radial 
velocities of probable NGC 6624 member stars. Ideally, about one third of the 
measurements should lie above the respective dispersion profile. Bottom row: 
maximum acceleration within the cluster along the line-of-sight at a given 
radius for the cluster models compared to the period derivatives of the three 
pulsars and 4U 1820-30. Only very cuspy density distributions can cause the 
observed Doppler shift of 4U 1820-30.\label{fig:models}}
\end{figure*}

\begin{figure*}
\includegraphics[width=1\textwidth]{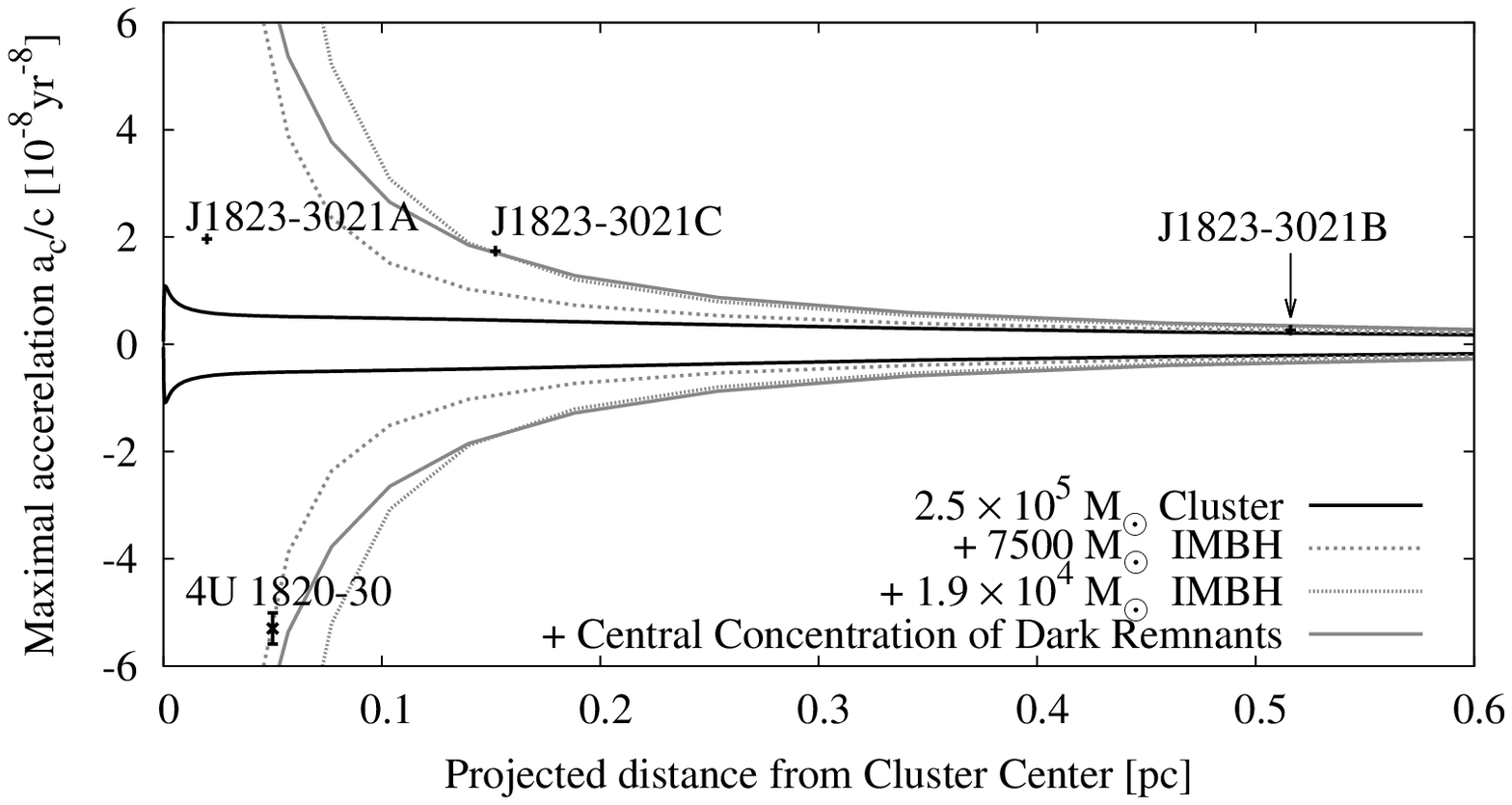}
\caption{Expected maximal cluster acceleration depending on the projected 
radial position in the cluster, based on the cluster model with 
$\unit[2.5\times10^{5}]{M_{\odot}}$ in which the mass profile follows the 
observed surface brightness profile (thick black line). Also shown is this 
model but with an additional IMBH of $\unit[7500]{M_{\odot}}$ (dashed line) or
$\unit[1.9\times10^{4}]{M_{\odot}}$ (dotted line), respectively. Alternatively,
we show a cluster model with a central concentration of dark remnants (grey 
line), i.e. for which the mass profile deviates from the light profile. A 
central slope of the mass profile of 1.7 can explain the observed period 
derivatives.
\label{fig:a_over_c_vs_d}}
\end{figure*}

Based on available observational constraints, we generated dynamical models of 
NGC 6624 to estimate the gravitational acceleration on 4U 1820-30 and on three 
further pulsars in NGC 6624, for which period derivatives have been measured 
\citep{2012ApJ...745..109L}. In Table \ref{tab:List-of-Pulsars} we list the 
relevant properties of these three pulsars together with the corresponding data
from 4U 1820-30. What is striking with these four targets is that their period 
derivatives have the same magnitude, which could be interpreted of being 
affected by the same gravitational pull\footnote{For an environment independent
explanation for the pulsar shifts see \citet{2012ApJ...745..109L}.}. Moreover, 
all four of them lie within about 0.5 pc (projected) radius of the cluster 
center. 

To create the models, we used the \textsc{McLuster}
code\footnote{\texttt{https://github.com/ahwkuepper/mcluster.git}}  
\citep{2011MNRAS.417.2300K}. The observational constraints were given through 
the pulsar data, an extended surface brightness profile 
\citep{2006AJ....132..447N}, and radial velocity measurements of 
\citet{1989AJ.....98..596P} and \citet{2011MNRAS.414.2690V}. 
 
\citet{2006AJ....132..447N} combined HST/WFPC2 photometry of the center of NGC 
6624 with ground-based wide-field data of \citet{1995AJ....109..218T} to create
a V-band surface brightness profile extending from 0.26 arcsec to 76 arcsec, 
which corresponds to 0.01--3.0 pc. We fit a cuspy Nuker profile 
\citep{1995AJ....110.2622L} to this surface brightness data by least squares 
minimization, where we used the uncertainties given by 
\citet{2006AJ....132..447N} as weights (thick black line in the upper 
row panels of Fig.~\ref{fig:models}). The Nuker surface density
profile has the form
\begin{equation}
I (R) = 2^{\left(\eta-\gamma\right)/\alpha} I_0\,\left(\frac{R_c}{R}\right)^\gamma\,\left( 1 + \left( \frac{R}{R_c}\right)^\alpha \right)^{\left(\gamma-\eta\right)/\alpha},
\end{equation}
where $I_0$ is a scale surface density, $\gamma$ and $\eta$ are two power-law 
slopes, and $R_c$ is the break radius, where the profile transits from one 
slope into the other. We set the transition parameter $\alpha = 2$. The 
resulting values were $\gamma = 0.31$ for the inner profile slope, 
$\eta = 1.85$ for the outer profile slope and a break radius of 
$R_c = 0.22$\,pc, similar to the values found by 
\citet{2006AJ....132..447N}, who used a King profile 
(\citealt{1966AJ.....71..276K}). To create a mass profile from that, we first 
deprojected the profile into 3D assuming spherical symmetry, and then created a
corresponding mass profile assuming that mass follows light in the cluster, 
i.e. that the cluster is not mass-segregated (in Section \ref{sec:ACCDR} we are
going to drop this assumption to show that the cluster core may harbor a core 
of dark remnants). We then scaled this profile to a given total mass within a 
radius of 30 pc, that is between the tidal radius estimated by 
\citet{1987ApJ...320..482L} of 26.8 pc and the estimate from the modeling of 
\citet{1989AJ.....98..596P} of about 33 pc. From this mass profile we got a 
velocity dispersion profile by solving the Jeans equation, assuming virial 
equilibrium and velocity isotropy. The latter seems to be appropriate, as 
\citet{1989AJ.....98..596P} find that radially anisotropic dynamical models 
give poorer fits to the data.

From this 3D velocity dispersion profile we computed the line-of-sight (LOS) 
velocity dispersion profile following \citet{1982MNRAS.200..361B}. We compared
the resulting LOS velocity dispersion profile of these models against 18 
DAO/CFHT radial velocities of probable NGC 6624 member stars from 
\citet{1989AJ.....98..596P} and 5 radial velocities measured by 
\citet{2011MNRAS.414.2690V} with high-resolution Keck/NIRSPEC data (middle row 
panels of Fig.~\ref{fig:models}). For this purpose we calculated projected 
distances from NGC 6624's center at RA = 18:23:40.51, Dec = -30:21:39.7 
\citep[J2000, ][]{2010AJ....140.1830G} to the coordinates given in 
\citet{2011MNRAS.414.2690V}. Their data yields a velocity dispersion of 
$\approx6$\,km/s between 0.2 pc and 1.0 pc projected distance from the center. 
For the \citet{1989AJ.....98..596P} radial velocity data we used the radial 
distances from the cluster center as stated in their Table 2, due to a lack of 
proper coordinates. Together their data sets give a systemic velocity of 
$v_{\scriptsize{\mbox{mean}}} = 53.5$\,km/s and a velocity dispersion of 
$\sigma\approx5.5$\,km/s within 1.3\,pc projected distance. For a good fitting 
model, about 68\% of the measured velocities should lie below the LOS velocity 
dispersion profile.

To scale the cluster light profile to a mass profile we need the mass-to-light 
ratio of the cluster population or its total mass. Unfortunately, the total 
mass of NGC 6624 is relatively unconstrained, as neither its total integrated 
magnitude nor its distance modulus are well measured. Hence, we can only 
collect and compare estimates based on various approaches: From multi-mass King
model fitting, \citet{1989AJ.....98..596P} find a mass between 
$100,000\,\mbox{M}_\odot$ and $220,000\,\mbox{M}_\odot$, and global 
mass-to-light ratios, $M/L_{V}$, between 1.9 and 2.3. The cluster's integrated 
absolute V-band magnitude is -7.49 mag \citep{1996AJ....112.1487H}, which 
translates into a cluster luminosity of 
$L/L_\odot = 10^{(M_{V,\odot}-M_V)/2.5} \approx85,000$. Thus, using a global 
mass-to-light ratio $M/L_{V}=1-3$  would yield a mass of 
$\unit[8.5\times10^{4}]{M_{\odot}}$ to $\unit[2.5\times10^{5}]{M_{\odot}}$. 
However, NGC 6624 is embedded in a rich foreground and background of bulge and
disk stars. In addition, the uncertainty in NGC 6624's distance has a strong 
influence on this estimate. If we just take the central surface brightness 
profile of \citet{2006AJ....132..447N}, which is less affected by contamination,
transform it into a luminosity density profile and scale it to the luminosity 
profile of our test models, we find that a model with 
$\unit[1.5\times10^{5}]{M_{\odot}}$ matches the data well if we assume a global
mass-to-light ratio of about 3. However, this estimate involves an 
extrapolation of the Noyola data from 3 to 30 pc, and also involves the 
uncertain cluster distance which significantly affects this mass estimate. As a
conclusion from these exercises we take that the mass of NGC 6624 should be 
somewhere between $\unit[1\times10^{5}]{M_{\odot}}$ and 
$\unit[2.5\times10^{5}]{M_{\odot}}$.

A mass of $M = 250,000$\,M$_{\odot}$ yields an estimated Jacobi radius, $r_J$, 
of $25-30$\,pc for NGC 6624. For this estimate we assume a circular Galactic 
orbit and a local circular velocity of between $170-220$\,km/s at the assumed 
Galactocentric radius, $R_{GC}$, of 1.2\,kpc, and we use
\begin{equation}
r_J = \left(\frac{GM}{2\Omega^2}\right)^{1/3},
\end{equation}
where $\Omega = V_c/R_{GC}$ is the local angular velocity 
\citep[e.g.][]{2010MNRAS.401..105K}. For this mass we get a line-of-sight 
velocity dispersion of $\sigma_{\scriptsize{\mbox{LOS}}}\approx6$ km/s within a
radius of $\unit[1.3]{pc}$ (central panel of Fig.~\ref{fig:models}). 
$M = 250,000$\,M$_{\odot}$ therefore seems to be a reasonable assumption for 
the mass of NGC 6624.

However, the maximal acceleration at the position of 4U 1820-30 of this model 
is $\frac{a_{\scriptsize{\mbox{c,max}}}}{c}\approx\unit[2.3\times10^{-9}]{yr^{-1}}$. 
This value is an order of magnitude too low, but also the maximal accelerations
for the three other pulsars are too low, as can be seen from the thick black 
line in the middle panel of the lower row of Fig.~\ref{fig:a_over_c_vs_d}. In 
this figure we have plotted the distance-dependent expected maximal 
acceleration from our dynamical model together with the position of the four 
targets. To test if a higher cluster mass could solve the problem, we scaled 
the \citet{2006AJ....132..447N} to the cluster's mass profile using an 
unrealistic global mass-to-light ratio of 10 (top panel in the left column of 
Fig.~\ref{fig:models}). In this case we get a cluster mass of 
$\unit[5\times10^{5}]{M_{\odot}}$ within a radius of 30\,pc. But even for this
high mass-to-light ratio (which might be biased through the uncertain distance 
modulus), the maximal acceleration at the position of 4U 1820-30 would only be 
$\frac{a_{\scriptsize{\mbox{c,max}}}}{c}\approx\unit[4.6\times10^{-9}]{yr^{-1}}$ 
(lower left panel of Fig.~\ref{fig:models}). The same applies to the maximal 
accelerations at the radii of the two inner pulsars. Higher cluster masses 
would make the observed velocity dispersion rather unlikely (see the thick 
black line in the middle left panel of Fig.~\ref{fig:models}). Apparently, 
the cluster profile as given through the surface brightness profile can neither
explain the acceleration of 4U 1820-30 nor of the other three pulsars. This 
suggests that there is either an additional, significant source of gravity 
close to these objects or that the mass-to-light ratio of NGC 6624 is 
significantly higher in the center than in the outer parts. In the following 
we discuss these possible scenarios: A stellar mass dark remnant, an IMBH or a 
central concentration of dark remnants.

\subsection{Stellar mass dark remnant}\label{sec:FlyBy}

The negative observed time period might be explained by the radial 
gravitational acceleration towards a nearby, so far undetected, dark remnant. 
This dark object, passing close to 4U 1820-30, can be a white dwarf, a neutron 
star or a stellar mass black hole. In order to constrain the allowed parameter 
range, we performed computations with the $N$-body software \textsc{catena} 
(\citealp{2006MNRAS.373..295P}). In this way we investigated the dynamical 
stability of the 4U 1820-30 system which we used as the main criterion. 4U 
1820-30 was set up as documented by \citet{2012ApJ...747....4P}, with a third 
main sequence star in a narrow orbit around the inner WD-NS binary. The 
additional companion scenario is the only theory which can explain the 171 day 
period so far.

\begin{figure}
\includegraphics[width=1\columnwidth]{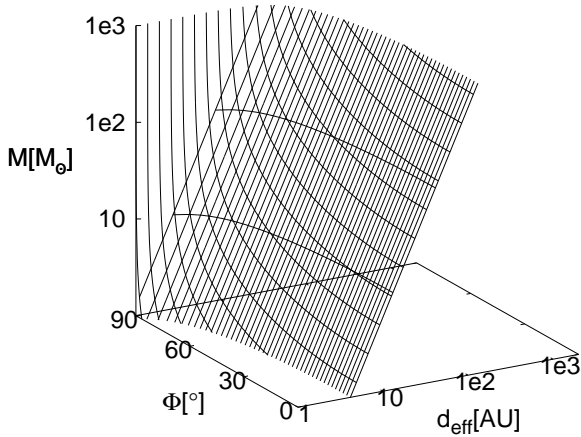}
\caption{For a given line-of-sight inclination $\Phi$ and dark remnant mass
$\unit[]{M}$, the plot illustrates the required distance 
$d_{\scriptsize{\mbox{eff}}}$ to achieve the gravitational acceleration.
\label{fig:bhplot}}
\end{figure}

We then tested the dynamical stability of this triple in the gravitational 
field of an external perturber, i.e. a forth object. The mass range of this 
dark remnant was varied between $\unit[1-100]{M_{\odot}}$. Depending on the 
actual mass of the perturber, the required distance 
$d_{\scriptsize{\mbox{eff}}}$ to 4U 1820-30 (i.e. center of mass) was 
calculated with the help of Eq.~\ref{eq:Fig6} by assuming a radial line of 
sight acceleration. Fig.~\ref{fig:bhplot} illustrates the dependency of a 
non-zero line-of-sight inclination.
 
The orientation of the velocity vector of the dark remnant with respect to the 
triple was randomly generated. The only boundary condition was that without 
gravitational interaction the dark remnant would pass 4U 1820-30 with a minimal
distance of $d_{\scriptsize{\mbox{eff}}}$. The velocity of the remnant was 
drawn from a normal distribution around the 3D velocity dispersion of the 
cluster ($\unit[6]{km/s}$). The initial distance of the perturber was more than
10 times the distance $d_{\scriptsize{\mbox{eff}}}$. In order to increase the 
statistical significance, this procedure was repeated ten times for every 
considered perturber mass\footnote{The whole computational timescale was of the
order of 1.5 month on a conventional quad-core CPU.}.

\begin{figure}
\includegraphics[width=1\columnwidth]{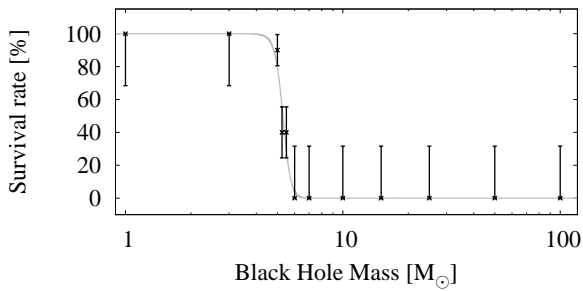}
\caption{Survival rate of 4U 1820-30 after a flyby of a dark remnant of 
different masses. The solid line shows an exponential fit of the form 
$f(x)=1/\left(1+\exp\left(a\cdot x+b\right)\right)$ 
through the data. Error bars for mass ranges $\unit[1 - 3,6 - 100]{M_{\odot}}$
are Poisson errors, while for the mass range $\unit[3-6]{M_{\odot}}$ the errors
were determined using bootstrapping. \label{fig:Res-Sim}}
\end{figure}

\begin{figure}[h!]
\includegraphics[width=1\columnwidth]{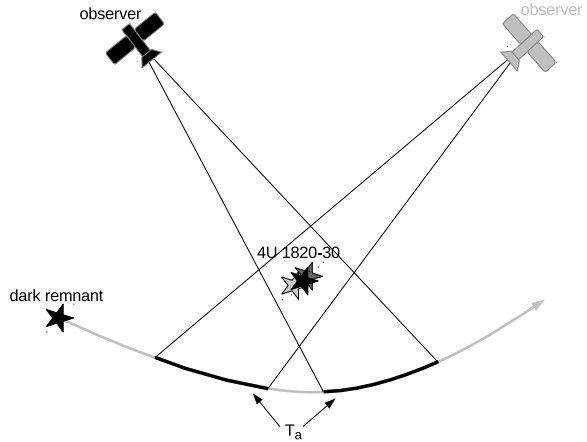}
\caption{Illustration of a typical flyby scenario. $T_{a}$ is the time where 
the acceleration is in the $5\sigma$-region of our measured value. Note that 
there exist two periods where independent observers could measure the effect.
\label{fig:FlyBy}}
\end{figure}

The results are shown in Fig.~\ref{fig:Res-Sim}, were we plotted the fraction 
of cases where the triple remains intact i.e. the survival rate against the 
dark remnant mass. It is apparent that for masses greater than 
$\unit[6]{M_{\odot}}$, the LMXB would not survive such a flyby, leaving us with
relatively light remnants. Analogous simulations with a dark remnant in a bound 
orbit yield comparable results. 

Light remnants must be close to the 4U 1820-30 system in order to generate the
required acceleration. However, the time they can maintain the required 
acceleration is strongly reduced. The timespan $T_{a}$, where the acceleration 
is in the $5\sigma$-region of our measured value, varied between 
$\unit[\sim20]{yr}$ (for a $\unit[1]{M_{\odot}}$ dark remnant) to
$\unit[\sim160]{yr}$ (for a $\unit[5.5]{M_{\odot}}$ dark remnant), which is 
illustrated in Fig.~\ref{fig:FlyBy}. These numbers give a rather quantitative 
than qualitative result, but they indicate that the chance to see a flyby
right now is considerably low. 

Hence, the flyby scenario is a rather short lived phenomenon, which can not be 
excluded yet, but by measuring the second derivative, $\ddot{P}$, of the period,
one could in fact constrain it. However, the chance that the other three 
pulsars are also experiencing a flyby event is highly unlikely. 

\subsection{IMBH}

\citet{2013A&A...552A..49L} present evidence for the existence of intermediate 
mass black holes (IMBHs) in galactic globular clusters. At first glance such an
object would be a considerable solution; the negative period derivative of 4U 
1820-30 could be explained by the radial gravitational acceleration 
(Eq.~\ref{eq:Fig6}) towards an IMBH of at least $\unit[7500]{M_{\odot}}$ (see 
Fig.~\ref{fig:a_over_c_vs_d}).

The idea of NGC 6624 having an IMBH in its center is not a new one: already 
\citet{1976ApJ...204L..83B} suspected that NGC 6624 could be harboring a 
massive black hole. 

However, there are certain arguments which contradict this idea: First of all, 
the period shifts of the three pulsars can only be explained by the 
$1/r$-potential of a very massive IMBH (Fig.~\ref{fig:a_over_c_vs_d}) in excess
of $\unit[19000]{M_{\odot}}$.

Furthermore, assuming the triple nature for 4U 1820-30 as proposed by 
\citet{2012ApJ...747....4P}, the most stringent argument against an IMBH 
causing the negative period derivative is related to its stability. In the 
previous subsection, we analyzed the stability of the 4U 1820-30 system. We 
found that massive objects in excess of $\unit[6]{M_{\odot}}$ would disrupt the
triple (as can be seen in Fig.~\ref{fig:Res-Sim}). However, 
the IMBH would have a mass at least 3000 times greater than that limit. This 
would imply that the period derivative we observe for 4U 1820-30 is a unique 
event with a short duration as the triple would be destroyed by tidal forces 
rather quickly.

\subsection{A Central Concentration of Dark Remnants}\label{sec:ACCDR}

As an alternative to the above scenarios, we may consider that the assumption 
of a constant mass-to-light ratio across the cluster, which we used for 
generating our dynamical models, is not fulfilled. That is, the surface 
brightness profile may not be a good tracer of the mass distribution. The 
central mass-to-light ratio may be significantly enhanced by dark remnants, 
such as white dwarfs, neutron stars and/or stellar mass black holes, which sank
into the center via mass segregation. \citet{1969ApJ...158L.139S} showed that 
heavier objects like black holes can form a dynamically decoupled subsystem 
within a star cluster, if their total mass is a significant fraction of the 
total cluster mass. In star clusters that contain a high fraction of massive 
dark remnants, the result of such a Spitzer instability is a high central 
concentration of dark mass, which inflates the central mass-to-light ratio (see 
\citealp{2011ApJ...741L..12B} for the extreme cases of \textit{dark star 
clusters}). Using idealized two-component systems, \citet{2013MNRAS.432.2779B} 
showed that the lifetime of such a dynamical subsystem is longer the smaller 
the difference in the masses of the heavier objects and the lighter objects 
are. That is, a subsystem of many lower-mass dark remnants, such as neutron 
stars or white dwarfs, would survive for a longer time than a subsystem of 
black holes with masses of several solar masses, and would therefore be more 
likely to have survived until the present day within an old globular cluster 
like NGC 6624. 

We may assume that such a dark remnant concentration has formed inside the 0.22 
pc break radius of the radial surface brightness profile, and that therefore 
inside this radius the surface brightness profile is flatter than the mass 
density profile. To see the effect of such a dark remnant cusp inside NGC 6624,
we generated models with deviating power-law slopes inside the break radius, 
which is equivalent to having a radially increasing mass-to-light ratio towards 
the center. The surface brightness profile shows an inner slope of 0.31, we 
varied this slope in steps of 0.1 up to the outer profile slope of 1.85. The 
resulting mass density profiles, line-of-sight velocity dispersion profiles and
maximum acceleration profiles for all three model masses are shown in 
Fig.~\ref{fig:models}.

When we include a high amount of dark mass in the center of our model of NGC 
6624 with a mass of 250,000\,M$_{\odot}$, and assume that the mass density 
profile within the break radius has a slope of about 1.7 instead of the 0.31 
observed for the surface brightness profile, all three period shifts of the 
inner pulsars can be explained by the same phenomenon, invoking less of a 
coincidence than the other scenarios. The amount of (dark) mass necessary would
be of the order of 70,000\,M$_\odot$ and would be limited to within the break 
radius of 0.22 pc. Interestingly, even our model with a mass of 
150,000\,M$_{\odot}$ could nearly reproduce the pulsar Doppler shifts if the 
mass density profile followed the same power-law slope of $\approx1.8$ inside 
the break radius like outside the break radius. This would correspond to a dark
remnant mass of about 55,000\,M$_\odot$ inside the break radius. 

Both configurations would imply that about $27\%-37\%$ of the cluster mass is 
locked up in dark remnants inside the cluster's break radius, yielding a dark 
remnant density of about $10^6$\,M$_\odot$/pc$^3$. This significant 
contribution from dark remnants has to be in accordance with stellar evolution
models. To check this we used \textsc{McLuster} for generating aged stellar 
populations. If we assume that NGC 6624 was born with a regular initial mass 
function (e.g. \citealp{2001MNRAS.322..231K}), and neglect dynamical evolution 
for a crude estimate, the dark remnant mass fraction at present day, i.e. after
12 Gyr of stellar evolution, would be $30\%$. This is in good agreement with 
our findings. Most of this mass would come from white dwarfs ($21\%$), and 
$9\%$ of the total mass would be in neutron stars and black holes. However, 
dynamical evolution (cluster mass loss, preferential loss of low-mass stars, 
and ejections from the core as a consequence of few-body interactions), as well
as birth kicks for neutron stars and black holes, will alter the contribution 
from dark remnants to the total cluster mass over time. Based on its 
present-day mass function, \citet{2012MNRAS.422.2246M} suggested that 
NGC\,6624 may have been born with a top-heavy IMF, which would certainly 
help to produce such a dark-remnant enriched cluster. Performing detailed 
$N$-body computations of NGC 6624 to show how a star cluster can form such a 
dark-remnant feature would be desirable, but is beyond the scope of this paper.

The dark mass concentration would have consequences for the radial velocity 
dispersion of stars within the central cusp of NGC 6624: The cuspy density 
profiles show an enhanced peak LOS velocity dispersion inside the break radius 
(gray lines in the middle row panels in Fig.~\ref{fig:models}). Interestingly, 
\citet{1992A&A...258..302Z} find with integrated spectroscopy of the inner 
$\approx$0.16\,pc that the velocity dispersion of NGC 6624 in this part is of 
the order of 9\,km/s. In contrast to that, most radial velocity measurements 
for NGC 6624 were outside this radius where the effect of the central density 
cusp is weaker. For a central slope of 1.7 and a cluster mass of 
250,000\,M$_\odot$ 4 out of 23 stars would still lie above the LOS velocity 
dispersion profile. For a slope of 1.8 and a mass of 150,000\,M$_\odot$ 7 out 
of 23, or 30\%, could lie outside the predicted profile. The peak dispersion of
this model is $\approx8.5$\,km/s. We therefore conclude that this model 
reflects the current data reasonably well.

Interestingly, a similar configuration of a dark remnant concentration has been
discussed by \citet{2008ApJ...676.1008N} for the center of the Milky Way 
globular cluster $\omega$\,Centauri. They estimated that $10^4$ dark remnants 
would be necessary within a radius of 0.05 pc of $\omega$\,Centauri's core to 
reproduce the observations. In the end, they favored an IMBH as the explanation 
for the observed surface density and velocity dispersion profile of this 
cluster. They argued that such a high concentration of dark remnants could not 
have been able to form in $\omega$\,Centauri as its present-day half-mass 
relaxation time is so long. However, the authors did not take into account that
a Spitzer instability significantly speeds up the formation of such a dark 
remnant core. For 47 Tuc, which has a nearly identical 
metallicity and age as NGC 6624 \citep{2000AJ....120..879H}, 
\citet{2011MNRAS.410.2698G} find that dark remnants make up about 34\% of 47 
Tuc's total mass and that the central dark remnant fraction is even larger than 
$50\%$. \citet{1995AJ....109..209G} came to similar conclusions fitting Jeans 
models to the much better and more abundant observational data for 47 Tuc. 

The idea that NGC 6624 could harbor a high amount of dark remnants has also 
been addressed by \citet{1992ApJ...392...86G}, who used Fokker-Planck models to
analyze the dynamical state of NGC 6624. Following an independent approach from 
ours, their resulting model had a total mass of 
$M_{tot}=\unit[1.5\times10^{5}]{M_{\odot}}$, in which about 37\% of the total 
mass, or $M_{dark}=\unit[5.6\times10^{4}]{M_{\odot}}$, was bound in dark 
remnants. If NGC 6624 had been able to keep a fair amount of its white dwarfs, 
black holes and neutron stars in the past, we may in fact be observing the 
result of a Spitzer instability in the center of NGC 6624. The further 
investigation of this cluster is therefore of fundamental interest and may 
even yield valuable insights into the nature of dark remnant formation.

\section{Conclusion\label{sec:Conclusion}}

By reducing 16 years of RXTE data (Section 
\ref{sec:Data,-Data-Reduction-Analysis}), we solved one important question 
concerning the, in many aspects extreme, low-mass X-ray binary 4U 1820-30: Its 
period derivative is indeed negative with a value of 
$\dot{P}/P=\unit[\left(-5.3\pm0.3\right)\times10^{-8}]{yr^{-1}}$. However, by 
doing so, we created a new puzzling question, as this is completely contrary to 
the expectations for a neutron star accreting mass from a Roche lobe-filling 
white dwarf in an isolated binary system.

Fortunately, the 4U 1820-30 system is not in isolation but in the very center 
of the Milky Way globular cluster NGC 6624. We therefore present some possible 
interpretations of this observation in Section \ref{sec:Interpretation}. We 
argue that this period shift is primarily due to gravitational acceleration 
from a non-luminous object. This could either be an intermediate-mass black 
hole (IMBH) in the center of NGC 6624, a nearby dark stellar remnant on a 
flyby, or a central dark remnant concentration within NGC 6624. 

We showed that an IMBH and a heavy dark remnant would disrupt the complex 4U 
1820-30 system, so that its negative period derivative would purely be a chance 
observation. The same could be the case for a lighter dark remnant, since the 
time scales are quite short. However, if three more pulsars within NGC 6624 are 
considered for the interpretation of the 4U 1820-30 observations, it appears 
more likely that all four of them are being strongly accelerated within the 
gravitational field of NGC 6624. In order to explain all four period 
derivatives, an extended concentration of non-luminous mass within the center 
of NGC 6624 would be necessary. We therefore suggest that a dark subsystem of 
black holes, neutron stars and/or white dwarfs may have formed via Spitzer 
instability in NGC 6624. Thus, a more detailed study of NGC 6624's dynamical 
state could yield valuable insights into the formation mechanisms of dark 
remnants.

A better understanding of the internal processes of LMXBs may explain the 
negative period derivative 4U 1820-30 without the need for a Doppler shift from 
NGC 6624 (e.g. \citealp{1987SvAL...13..122P}, \citealp{2002ApJ...565.1107P}). 
However, as we have shown here, the presence of additional matter in the 
cluster core can explain not just 4U 1820-30's period derivative, but also the 
strong Doppler shift observed for the pulsars at larger radii.

To shed more light on the problem, it would be helpful if other X-ray satellite 
missions could continue to observe 4U 1820-30. The RXTE satellite ceased 
science operations on 3 January 2012. As presented in Section 
\ref{sec:Data,-Data-Reduction-Analysis}, the data we used had quite a coarse 
temporal resolution and most of the observations were not designed to observe 
the 685 sec periodicity. We hope that further observations will not only reduce 
the error on the period derivative but also help to evaluate the second period 
derivative. This value would allow to probe if the negative period derivative 
is a local transient phenomenon or if, as the pulsar period derivatives seem to 
indicate, is due to concentration of some sort of non-luminous mass in NGC 
6624's center. Also, the determination of the period derivative of so far 
unconstrained pulsars (J1823\textminus{}3021D,E,F) would give further insights 
to the structure of NGC 6624. Especially, a measurement of the period 
derivative of pulsar J1823\textminus{}3021D would be important, as it lies at 
the same (projected) distance from the center of NGC 6624 as 
J1823\textminus{}3021C does. Another possibility to improve our 
knowledge of the dark component of NGC 6624 could be the search and analysis of
additional short-period LMXBs like the recently discovered COM\_Star1 
\citep{2014ApJ...784L..29D}.

We thank Snezana Prodan for valubale discussion during her visit of the SPODYR 
research group in Bonn. We would also like to thank Jay Strader and the referee
for helpful comments. MB would like to acknowledge support through DFG grant KR 
1635/39-1. AHWK would like to acknowledge support through DFG Research 
Fellowship KU 3109/1-1 and from NASA through Hubble Fellowship grant 
HST-HF-51323.01-A awarded by the Space Telescope Science Institute, which is 
operated by the Association of Universities for Research in Astronomy, Inc., 
for NASA, under contract NAS 5-26555.

Results provided by the ASM/RXTE teams at MIT and at the RXTE SOF and GOF at 
NASA's GSFC.

\bibliographystyle{apj}
\bibliography{Ref}

\end{document}